\documentclass[reprint,superscriptaddress,amsmath,amssymb,prl,aps,noeprint]{revtex4-2}
\usepackage{graphicx}
\usepackage{dcolumn}
\usepackage{bm}
\usepackage{braket}
\usepackage{hyperref}
\hypersetup{
    colorlinks=true,
    linkcolor=blue,     
    urlcolor=blue,
    citecolor=blue,
}
\usepackage[capitalise]{cleveref}
\usepackage{multirow}
\usepackage{tabularx}
\usepackage[euler]{textgreek}
\usepackage{xcolor}
\usepackage[version=4]{mhchem}
\usepackage{float}
\usepackage[mathlines]{lineno}
\usepackage{comment}
\usepackage{enumitem}
\usepackage{xparse}
\usepackage{ifthen}
\usepackage{ulem}

\newboolean{draft}
\setboolean{draft}{True}
\def\NVm{\ensuremath{\mathrm{NV}^-}}
\NewDocumentCommand\change{om}{%
  \ifthenelse{\boolean{draft}}
  {\IfNoValueTF{#1}{{\color{orange}#2}}{{\color{lightgray}#1}/{\color{orange}#2}}}
  {#2}%
}
\def\tA2{^{3}\!A_{2}}
\def\tE{^{3}\!E}

\begin{document}

\title{Optical Stability and Photophysics of NV Centers in Diamond up to 120~GPa}

\author{Kin On Ho}
\thanks{These authors contributed equally to this work.}
\author{Cassandra Dailledouze}
\thanks{These authors contributed equally to this work.}
\affiliation{
	Université Paris-Saclay, CNRS, ENS Paris-Saclay, CentraleSupelec, LuMIn, F-91190 Gif-sur-Yvette, France
}
\author{Vytautas Žalandauskas}
\affiliation{
	Center for Physical Sciences and Technology (FTMC), Vilnius LT–10257, Lithuania
}
\author{Grégoire Le Caruyer}
\affiliation{
	Université Paris-Saclay, CNRS, ENS Paris-Saclay, CentraleSupelec, LuMIn, F-91190 Gif-sur-Yvette, France
}
\author{Marek Maciaszek}
\affiliation{
	Center for Physical Sciences and Technology (FTMC), Vilnius LT–10257, Lithuania
}
\affiliation{
	Faculty of Physics, Warsaw University of Technology, Koszykowa 75, 00-662 Warsaw, Poland
}
\author{Claire Roussy}
\author{Marie-Pierre Adam}
\author{Martin Schmidt}
\affiliation{
	Université Paris-Saclay, CNRS, ENS Paris-Saclay, CentraleSupelec, LuMIn, F-91190 Gif-sur-Yvette, France
}
\author{Loïc Toraille}
\author{Paul Loubeyre}
\affiliation{CEA DAM, DIF, F-91297 Arpajon, France}
\affiliation{Université Paris-Saclay, CEA, Laboratoire Matière en Conditions Extrêmes, 91680 Bruyères-le-Châtel, France}
\author{Lukas Razinkovas}
\email{lukas.razinkovas@ftmc.lt}
\affiliation{
	Center for Physical Sciences and Technology (FTMC), Vilnius LT–10257, Lithuania
}
\author{Jean-François Roch}
\email{jean-francois.roch@ens-paris-saclay.fr}
\affiliation{
	Université Paris-Saclay, CNRS, ENS Paris-Saclay, CentraleSupelec, LuMIn, F-91190 Gif-sur-Yvette, France
}

\date{\today}

\begin{abstract}
  The nitrogen vacancy (NV) center has emerged as a powerful quantum sensor in high-pressure research, with the observation of optically detected magnetic resonance at megabar pressures. However, some aspects of NV physics require further investigation to optimize the development of NV-based sensing under pressure. Here, we study both experimentally and theoretically the optical properties of the NV center under hydrostatic pressure. We investigate the evolution of the zero-phonon line (ZPL) position, radiative lifetimes, optical lineshapes, and photoionization thresholds of the NV center under pressures up to $\sim120~\mathrm{GPa}$. We also provide spectroscopic guidelines for performing high-pressure optical experiments. Our results confirm that the NV center remains a robust quantum sensor under extreme hydrostatic pressures, especially for magnetic characterization.
\end{abstract}

\maketitle

\textit{Introduction---}High-pressure applications of the negatively charged nitrogen vacancy (NV) center have attracted considerable attention in recent years due to their natural compatibility with diamond anvil cells (DAC) and robustness of its electronic spin properties under pressure~\mbox{\cite{Doherty2014Electronic,Ivady2014Pressure,Ho2021Recent}}. By utilizing optically detected magnetic resonance (ODMR) spectroscopy, the NV centers have become state-of-the-art sensors for studying both classical and quantum phase transitions under pressure~\mbox{\cite{Toraille2020Combined, Ho2020Probing, Yip2019Measuring, Lesik2019Magnetic, Hsieh2019Imaging, Dailledouze2025Imaging, Bhattacharyya2024Imaging, Mandyam2026Uncovering}}. %

Despite this progress, most high-pressure studies have primarily focused on spin-based sensing applications, while comparatively little attention has been given to other aspects of NV physics under pressure~\mbox{\cite{Ho2022Probing,Wang2021ac,Dai2022Optically,Ho2023Spectroscopic}}. In particular, photophysical studies under pressure have mainly focused on tracking the zero-phonon line (ZPL) shift~\mbox{\cite{Hilberer2023Enabling,Ho2023Spectroscopic,Doherty2014Temperature, Kobayashi1993High}}, while the pressure dependence of other important optical properties remains largely unexplored. As a result, the photophysical information, which is important to design and interpret high-pressure experiments, such as optimal excitation and photoluminescence (PL) collection conditions and optical stability, is still lacking.

To reach a megabar regime with NV centers, two distinct paths have been pursued by research groups. The first one consists of using (111)-oriented anvils, where the uniaxial stress applied on the back of the diamond is parallel to one NV orientation, which is exclusively used for ODMR detection~\mbox{\cite{Bhattacharyya2024Imaging, Wang2024Imaging, Mai2025,Hao2025,Chen2025, Mandyam2026Uncovering, Liu2025Evidence, Wen2025Imaging}}. On the other hand, near-hydrostatic conditions for the four NV orientations can be approached by using nanodiamonds inside the pressure-transmitting medium (PTM)~\mbox{\cite{Dai2022Optically}} or by microstructuring the tip of a (100)-oriented anvil~\mbox{\cite{Hilberer2023Enabling}}.

In this Letter, we present a comprehensive study of the pressure dependence of NV optical characteristics, emission and absorption lineshapes, photoionization thresholds, and radiative lifetimes under near-hydrostatic conditions. We begin with a density functional theory (DFT) investigation of NV optical properties under hydrostatic pressures up to 120~GPa. Experimentally, we examine the near-hydrostatic pressure dependence of the full NV PL spectrum under these extreme conditions. Experimental and theoretical results show a nearly pressure-independent emission brightness, together with an excellent agreement in the ZPL energies and overall PL lineshapes. Our results establish the NV center as a robust sensor under extreme pressures and provide spectroscopic guidelines for future high-pressure experiments.

\textit{Theory---}The NV center in diamond forms when a substitutional nitrogen atom is positioned adjacent to a carbon vacancy, forming a defect with $C_{3v}$ point-group symmetry~\cite{Doherty2013The}. In its negatively charged state, the electronic ground state is a spin triplet of $A_{2}$ orbital symmetry ($\tA2$). Optical excitation promotes the system to the triplet $\tE$ state, giving rise to a visible transition with a ZPL at 1.945~eV (637~nm). Between these levels lie intermediate spin-singlet states $^{1}\!E$ and $^{1}\!A_{1}$, which mediate spin-dependent intersystem crossing. This nonradiative pathway enables spin polarization and provides optical contrast for spin readout, making the NV center a versatile platform for quantum technology applications~\mbox{\cite{Rondin2014Magnetometry,Barry2020Sensitivity,Doherty2013The}}.

\begin{figure*}
	\includegraphics[width=\textwidth]{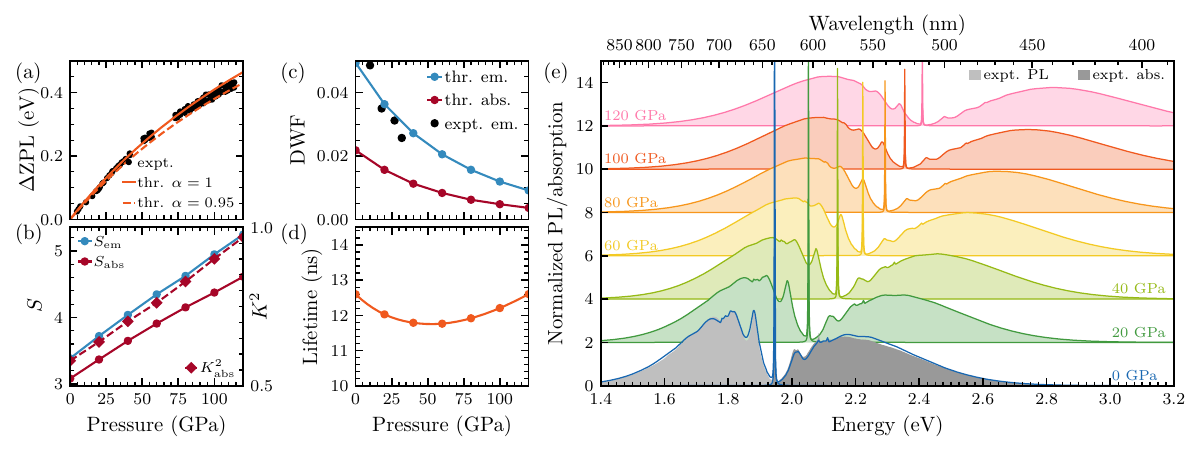}
	\caption{%
    (a) ZPL shift relative to ambient pressure. Black circles: experiment;
    orange solid line: theory (hydrostatic); orange dashed line: theory with near-hydrostaticity
    stress ($\alpha=0.95$).
    (b) Pressure-induced evolution of electron--phonon coupling parameters.
    Blue solid line: total HR factor $S$ (emission); red solid line: HR factor
    for JT-inactive modes (absorption); red dashed line: squared vibronic coupling
    constants $K^2$.
    (c) Pressure-induced evolution of Debye--Waller factors (DWF).
    Blue solid line: emission (theory); red solid line: absorption (theory); black circles: emission (experimental).
    (d) Pressure-induced evolution of theoretical radiative lifetime.
    (e) Theoretical PL and absorption lineshapes up to 120~GPa. The ZPL is
    aligned to experiment at ambient pressure and shifted at higher pressures by
    the theoretical rigid shift. Bottom lines: comparison with low-temperature
    spectra from Refs.~\cite{Kehayias2013Infrared} (PL, light gray)
    and~\cite{Manson2018} (absorption, dark gray). }
	\label{fig1}
\end{figure*}

\begin{figure}
  \includegraphics[width=\linewidth]{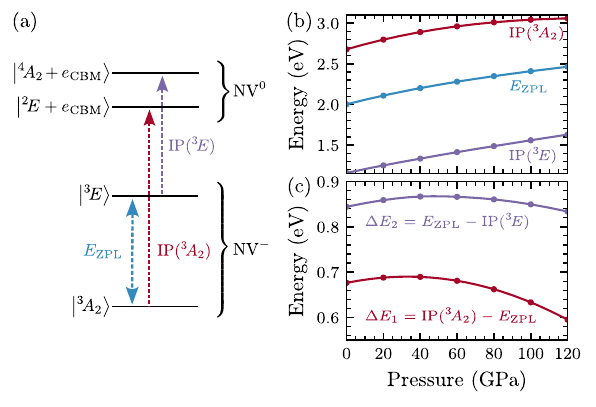}
  \caption{%
    (a) Photoionization of $\NVm$ from $\tA2$ and $\tE$
    states leading to $^2\!E$ and $^4\!A_2$ of NV$^0$.
    (b) Pressure evolution of ionization thresholds
    $\mathrm{IP}(\tA2)$ and $\mathrm{IP}(\tE)$ alongside
    with ZPL.
    (c) Ionization energies relative to ZPL.
    \label{fig:ionization}}
\end{figure}

At ambient pressure, the optical properties of the triplet transition of the NV center are well established~\mbox{\cite{Davies1976,Manson2018,Kehayias2013Infrared,Alkauskas2014First, Gali2019Ab,Razinkovas2021Vibrational}}. The PL spectrum consists of a sharp ZPL and a broad phonon sideband (PSB) extending to lower energies [see bottom left curve in Fig.~\ref{fig1}(e)], which roughly appears as a series of 65~meV phonon replicas associated with a vibrational resonance of the diamond lattice.
Experimental estimates of electron--phonon coupling provide a Huang--Rhys (HR) factor $S \approx 3.5\mbox{--}3.7$~\cite{davies1981a,Kehayias2013Infrared}, corresponding to the average number of phonons emitted during an optical transition~\cite{Alkauskas2014First,1950Huang}.
This corresponds to a Debye--Waller factor (DWF) of approximately 0.03, corresponding to the fraction of PL intensity contained in the ZPL. The radiative lifetime of this transition is experimentally estimated to be ${\sim} 12$~ns~\cite{Doherty2013The}. The absorption spectrum is not perfectly symmetric to PL [see bottom right curve in Fig.~\ref{fig1}(e)], reflecting the influence of the dynamical Jahn--Teller (JT) effect in the excited state as well as changes in vibrational structure upon excitation~\cite{Davies1976Optical,Razinkovas2021Vibrational}. Under optical illumination, the negatively charged state NV$^-$ is unstable, as an electron can be excited into the conduction band, converting it into the neutral state NV$^0$~\cite{aslam2013}.
This photoionization can occur either through direct one-photon ionization from the ground $\tA2$ state, or via a two-step mechanism in which the center is first excited to the $\tE$ state and subsequently ionized by a second photon~\cite{razinkovas2021a}, as illustrated in Fig.~\ref{fig:ionization}(a).

To investigate the optical properties theoretically, we employed spin-polarized DFT as implemented in the VASP code~\cite{kresse1996b}. Finite-pressure calculations were carried out by adjusting the lattice constant of the defective supercell according to the diamond equation of state~\cite{Occelli2003}. Ionization thresholds and transition dipole moments were calculated using the HSE06 hybrid functional~\cite{hse}, chosen for its reliable accuracy in reproducing the diamond band gap as well as the ZPL and radiative lifetime of the NV center. Excitation energies and phonon properties were computed with the computationally more efficient meta-GGA r$^2$SCAN functional~\cite{furness2020}, which provides good accuracy for both vibrational and electronic characteristics of the NV center~\cite{2023_Maciaszek_JCP}. Further computational details are given in Section~S3 of the Supplemental Material (SM).

For the description of electron--phonon interaction and optical lineshapes, we employed our previously developed methodology~\cite{Razinkovas2021Vibrational}. In brief, it employs an embedding approach to calculate vibrational structure of defects in the effectively dilute limit, overcoming the limitations of finite supercell models. Electron--phonon coupling parameters are obtained from the potential energy surfaces of the ground and excited states. For emission, the electron--phonon interaction is described by adiabatic HR theory~\cite{1950Huang}, which is valid since the final state ($\tA2$) can be treated adiabatically. In this case, mode-resolved HR factors $S_k$ are computed for all vibrational modes ($k$), and lineshapes are obtained via the generating-function approach~\cite{lax1952a,kubo1955}. In contrast, absorption requires explicit treatment of the JT effect. To this end, in addition to the HR factors for JT-inactive modes, for JT-active modes, we compute the squared dimensionless vibronic coupling constants $K_k^2$~\cite{obrien1980}, which play for linear JT coupling a role analogous to that of $S_k$ in HR theory~\cite{Razinkovas2021Vibrational,Zalandauskas2024Theory}. With these parameters at hand, we can account for the JT effect in the excited state to obtain the manifold of nonadiabatic vibronic states, from which the additional optical signatures arising from JT interactions can be estimated~\cite{Razinkovas2021Vibrational}. In the main text, we report only the aggregated quantities $S=\sum_k S_k$ and $K^2=\sum_k K_k^2$, which characterize the overall interaction strength. Further details, including the spectral decomposition of electron--phonon parameters and the calculation of optical lineshapes, are provided in Sec.~S4 of the~SM.

We begin by calculating the shift of the ZPL relative to ambient pressure, as shown in Fig.~\ref{fig1}(a) for pressures up to 120~GPa using r$^2$SCAN functional (see Sec.~S3 of the SM for a comparison with HSE06 under ideal hydrostatic conditions). The solid orange line corresponds to the case of perfect hydrostaticity, while the dashed orange line includes a small anisotropic component to account for deviations expected in the experiment (experimental details and anisotropic stress effects are discussed later). Experimental data points are presented for comparison, demonstrating excellent agreement with theory. We find a pronounced shift exceeding 400~meV at 120~GPa.

Next, we compute the pressure dependence of the parameters characterizing electron--phonon coupling, as shown in Fig.~\ref{fig1}(b). For emission, we plot the total HR factor $S$ (blue line). For absorption, we show the HR factor of symmetry-preserving JT-inactive modes (red solid line) together with the squared vibronic coupling constant $K^2$ (red dashed line, right axis). All parameters increase almost linearly with pressure. In particular, the PL HR factor rises from 3.39 at ambient pressure to 5.25 at 120~GPa, reflecting a pronounced enhancement of electron--phonon coupling. For absorption, the HR factor of JT-inactive modes increases from 3.08 to 4.61, while $K^2$ nearly doubles from 0.58 to 0.97, highlighting the strengthening of the JT interaction under pressure.

The strengthening of the electron--phonon coupling is directly manifested in the calculated spectra. With increasing pressure, the PSBs broaden and gradually lose their fine structure, an effect that is most pronounced in absorption. Figure~\ref{fig1}(e) presents the resulting emission (left curves) and absorption (right curves) lineshapes for pressures ranging from 0 to 120~GPa. For clarity, spectra are vertically offset in proportion to the applied pressure, with the bottom curves corresponding to ambient conditions. At ambient pressure, the theoretical spectra are compared with high-resolution, low-temperature measurements taken from Ref.~\cite{Kehayias2013Infrared} for PL and Ref.~\cite{Manson2018} for absorption. The agreement between theory and experiment is strong. For emission, the calculations capture both the individual spectral features and the redistribution of spectral weight across the PSB. For absorption, the characteristic double peak structure of the first phonon replica is reproduced, along with the overall spectral extent.

From the calculated optical lineshapes, we extract the DWF and its pressure dependence shown in Fig.~\ref{fig1}(c). At ambient pressure, the theory predicts values of 4.9\% for emission and 2.2\% for absorption. Both decrease monotonically with pressure, leading to more than a fivefold reduction at 120~GPa, where the DWF falls below 1\% for emission and reaches 0.36\% for absorption. Figure~\ref{fig1}(c) also shows approximate DWF values obtained from low-temperature (30~K) measurements of the emission spectrum up to 32~GPa (see Sec.~S2 of the SM for details of the extraction), displaying reasonable agreement with theory across this pressure range.

Figure~\ref{fig1}(d) presents the calculated radiative lifetime as a function of pressure. The dependence is nonmonotonic, reflecting the competition between the increase in ZPL energy and the reduction of the electric dipole moment of the transition under compression (see Sec.~S3 of the SM for details). However, the overall variation remains small (less than 1~ns) across the investigated pressure range.

Finally, we investigate the theoretical photoionization threshold of $\NVm$. As illustrated in Fig.~\ref{fig:ionization}(a), the system can be ionized from either the ground ($\tA2$) or the excited ($\tE$) states, with thresholds denoted $\mathrm{IP}(\tA2)$ and $\mathrm{IP}(\tE)$, respectively.
Ionization from the ground state leads to the $^2\!E$ ground state of NV$^0$, while ionization from the excited states results in the metastable $^4\!A_2$ state~\cite{razinkovas2021a}. Figure~\ref{fig:ionization}(b) illustrates the evolution of these thresholds alongside the ZPL as a function of pressure. At ambient pressure, we obtain $\mathrm{IP}(\tA2)=2.68$~eV and $\mathrm{IP}(\tE)=1.16$~eV, in agreement with previous reports~\cite{aslam2013,bourgeois2017,razinkovas2021a}. Both thresholds increase monotonically with the pressure, reaching 3.06 and 1.63~eV at 120~GPa, and follow a trend quantitatively similar to that of the ZPL. In Fig.~\ref{fig:ionization}(c), we plot the ionization thresholds relative to the ZPL energy. The ground-state ionization energy tracks the ZPL closely, while the deformation potential of $\mathrm{IP}(\tE)$ is somewhat smaller, though the difference is not substantial. These results suggest that under high pressure, if the excitation wavelength is adjusted to follow the pressure-induced shift of the ZPL, the charge-state stability of $\NVm$ should remain comparable to ambient conditions.

\begin{figure}
  \includegraphics[width=\linewidth]{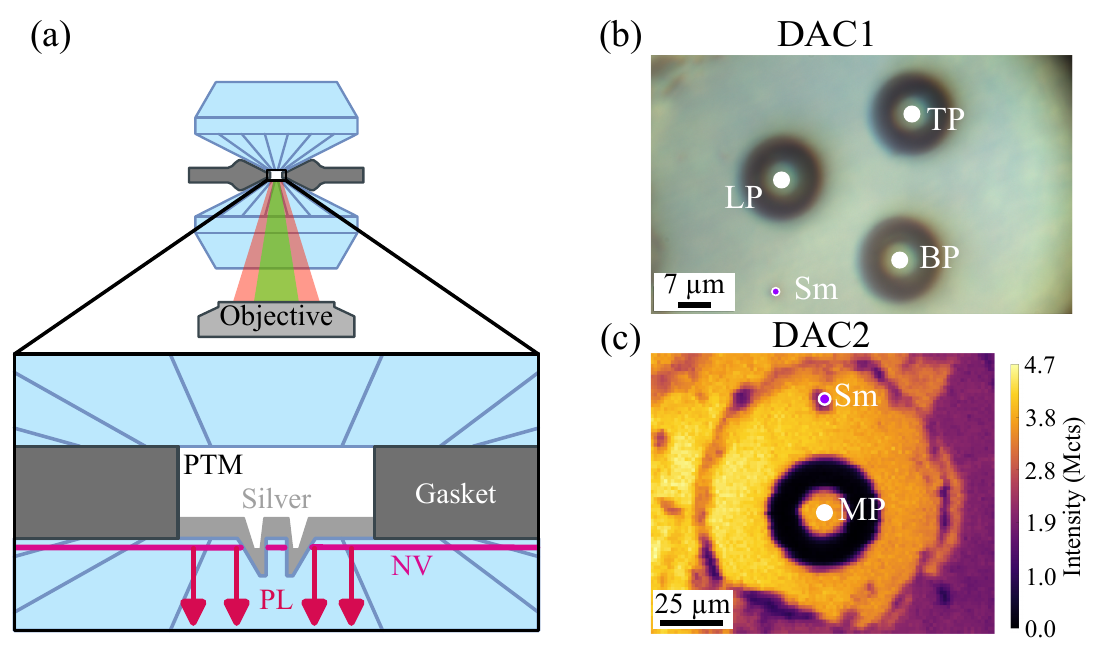}
  \caption{(a) Illustration of the DAC with a micropillar structure on (100)-diamond for reaching near-hydrostatic pressures~\cite{Hilberer2023Enabling}. We silver-coat the tip of the NV-implanted anvil to avoid interference from the top anvil and increase signal collection. (b) Image of DAC1 with three micropillars. (c) Confocal image of DAC2 with a single micropillar.\label{fig:setup}}
\end{figure}

\textit{Experiment---}To experimentally investigate the optical properties, we employ the DAC setup described in Ref.~\cite{Hilberer2023Enabling} and depicted in Fig.~\ref{fig:setup}(a). In this design, micropillars are machined on the anvil culet to reduce nonhydrostatic stress applied to the implanted NV layer. The deviation from hydrostatic conditions is quantified by the parameter $\alpha$, which describes the ratio between tangential and perpendicular stress components. In the micropillar geometry $\alpha \approx 0.95$, corresponding to near-hydrostatic conditions~\cite{Hilberer2023Enabling}. We performed two complementary experiments: (i) a single micropillar DAC [denoted MP, see Fig.~\ref{fig:setup}(c)] at $\approx 90$~K up to megabar pressures, and (ii) a three micropillar DAC [denoted TP, BP, and LP, see Fig.~\ref{fig:setup}(b)] at 30~K and moderate pressures. These two experiments allowed us to cover a wide range of experimental conditions in pressures and temperatures, thus reinforcing our findings. In both cases, the anvils were (100)-oriented and implanted with a near-surface, high-density NV layer following Ref.~\cite{Lesik2013Maskless}. Both DACs are loaded with argon as the PTM to provide near-hydrostatic stress~\cite{Klotz2009Hydrostatic}. We use samarium (\ce{SrB_{4}O_{7}\text{:}Sm^{2+}}) microcrystals~\cite{Datchi1997Improved} and the first-order diamond Raman~\cite{Akahama2004High} as the pressure gauge at low and high pressures, respectively.

\begin{figure}[t]
  \includegraphics[width=8.6cm]{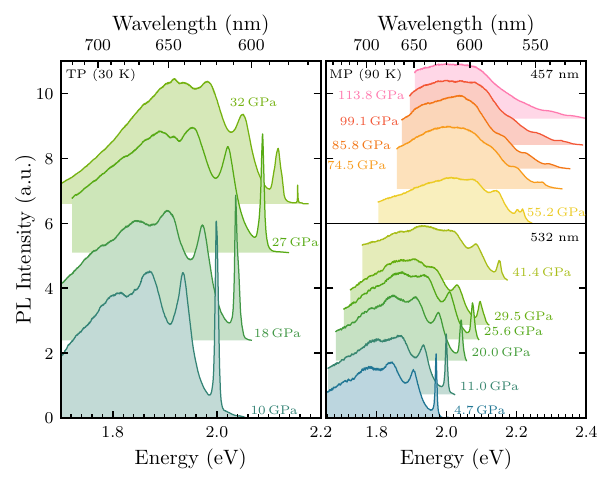}
  \caption{Pressure evolution of the experimental NV PL spectra. (Left) Measurements from the TP micropillar at 30~K using 532~nm excitation. (Right) Measurements from the MP micropillar at
    $\approx90$~K: bottom panel -- 532 nm excitation; top
    panel -- 457~nm excitation. Spectra in each panel are
    vertically offset by pressure-dependent scaling factors.\label{expt1}}
\end{figure}

The pressure evolution of the experimental PL spectra is shown in Fig.~\ref{expt1}. The left panel corresponds to data recorded from the TP micropillar at 30~K, while the right panel shows data from the MP micropillar measured at approximately 90~K. The two other micropillars from DAC1 (LP and BP) are showing very similar spectra so we will mainly show data from the TP micropillar.
In the latter, the bottom panel presents spectra recorded between 4.7 and 41.4~GPa using green laser excitation (532~nm), and the top panel shows those measured using blue excitation (457~nm) from 55.2 to 113.8~GPa. As predicted by theory, increasing pressure leads to a gradual decrease in the ZPL intensity and a broadening of the PSB, which becomes almost featureless above the megabar. As pressure increases, the ZPL exhibits a pronounced blue shift and additional broadening, which at high pressures evolves into a split ZPL due to the lifting of the $\tE$ state degeneracy under non-hydrostatic stress~\cite{Rogers2015Singlet}.
In Fig.~\ref{fig1}(a), we present the extracted ZPL energies alongside the theoretical predictions.  We calculated the ZPL position for perfect hydrostatic stress (orange solid line) and the lower branch of the splitted ZPL for $\alpha$ = 0.95 (orange dashed line). Unfortunately, modeling of the upper ZPL component remains challenging because of convergence issues arising from the close proximity of the split electronic states.
Nevertheless, modeling with $\alpha = 0.95$ reproduces the lower branch of the experimentally observed split ZPL with an excellent agreement. This validates the methodology of determining the stress-anisotropy parameter $\alpha$ from the magnetic field dependence of the ODMR spectra as done in Ref.~\cite{Hilberer2023Enabling}.

\begin{figure}[t]
	\includegraphics[width=\linewidth]{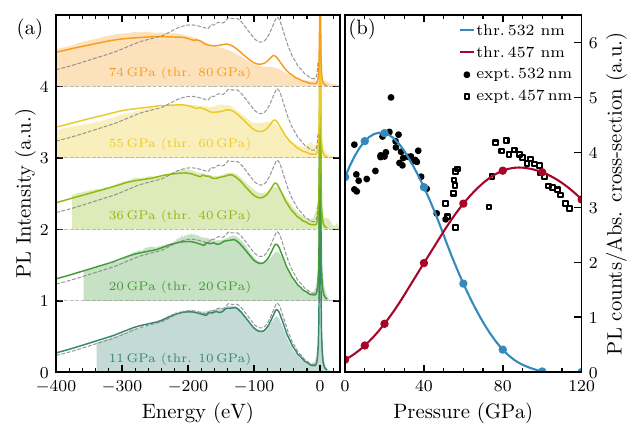}
	\caption{(a) Comparison between experimental and theoretical PSBs
    at approximately 10, 20, 40, 60, and 80~GPa. Solid lines and shaded areas represent calculations and experimental results. Dashed lines indicate the
    ambient-pressure reference. (b) PL photon yield (arbitrary units) as a
    function of pressure for excitation at 532~nm and 457~nm. Solid and open markers show the corresponding integrated experimental PL intensities, while solid blue and red lines represent the pressure-dependent absorption cross sections (arbitrary units) sampled at 532 and 457~nm from theory.}
	\label{expt2}
\end{figure}

In Fig.~\ref{expt2}(a), we compare the experimentally measured PL PSBs from selected MP micropillar measurements with the theoretical spectra calculated at the corresponding pressures. To enable a direct comparison of the lineshapes, all spectra are shifted so that the most prominent ZPL peak is aligned at zero energy. The pressure evolution of the PSB shows close agreement between theory and experiment, with a systematic transfer of spectral weight toward higher phonon energies as pressure increases. The remaining discrepancies are most pronounced at higher pressures and are likely attributable to ZPL splitting and broadening caused by residual non-hydrostatic stress and its spatial inhomogeneity, which affects the entire PSB.

From the spectral data, we can extract the total PL intensity as a function of pressure as shown in Fig.~\ref{expt2}(b). The two datasets, corresponding to the two excitation wavelengths 532~nm and 457~nm, form two broad domes that represent the pressure dependence of the absorption cross section at these excitation energies. For comparison, we overlay the absorption cross sections at 532~nm and 457~nm, obtained by sampling the calculated spectra at simulated pressures [Fig.~\ref{fig1}(e)] and interpolating between them. Since the excitation laser power and camera integration time were kept constant, and apart from the cross section difference, we can state that the NV quantum efficiency remains stable up to 120~GPa. This observation supports the calculations of the radiative lifetime presented in Figure~\ref{fig1}(d).


Taken together, these results indicate that increasing pressure enhances the electron–phonon interaction without significantly affecting the overall radiative efficiency of the NV center. Although the increase in electron-phonon interaction reduces the intensity of the ZPL and shifts the spectral weight of the PSB to higher energies, it does not impose fundamental limits on ODMR-based sensing, except for the need to adjust the excitation and detection wavelengths. This demonstrates that the optical readout mechanism remains robust under near-hydrostatic compression.
Recent works in (111)-oriented anvils have revealed the robustness of (111)-NV axis in non-hydrostatic conditions attributed to the surprising NV inter-system crossing at megabar pressures~\cite{Huang2025, Liu2026Strain}.
Together with recent demonstrations of NV-based magnetometry under extreme pressure~\cite{Hilberer2023Enabling, Dai2022Optically, Bhattacharyya2024Imaging,Wang2024Imaging,Mai2025,Hao2025, Chen2025}, these findings establish a consistent picture of the magnetic and optical stability of NV centers in diamond. They also encourage the use of NV-based sensing to explore magnetic phenomena in high-pressure (quantum) materials, including superhydrides, superionic ice, and metallic hydrogen if recoverable in the 100~GPa range.

\textit{Conclusion---}We present a thorough experimental and theoretical investigation of the optical properties of the NV center in diamond under extreme pressures. The strong agreement between theoretical and experimental results validates both methodologies and provides a consistent microscopic understanding of how pressure affects the optical response. Our findings establish the NV center as a robust magnetic-field sensor under extreme pressure and offer unambiguous spectroscopic guidelines for optimal excitation and photoluminescence collection.

\textit{Acknowledgements---}We thank Liam Hanlon for his help on the very first experiments and Florent
Occelli for the technical support.
This work has been supported by
the Region Ile-de-France in the framework of the DIM QuanTIP,
the ANR with the ESR/EquipEx+ e-diamant grant (ANR-21-ESRE-0031),
the SADAHPT grant (ANR-19-CE30-0027-01) and
the SENSEXTREME grant (ANR-22-QUA2-0009-04).
JFR acknowledges support from the Institut Universitaire
de France. VŽ and LR were supported by the QuantERA grant SensExtreme, funded by the Lithuanian Research Council (grant number S-QUANTERA-22-1).

\bibliography{references}

\end{document}